# Lorentz-boosted evanescent waves

Konstantin Y. Bliokh[1,2]

[1]*Nonlinear Physics Centre, RSPE, The Australian National University, Canberra, Australia*
[2]*Cluster for Pioneering Research, RIKEN, Wako-shi, Saitama 351-0198, Japan*

Polarization, spin, and helicity are important properties of electromagnetic waves. It is commonly believed that helicity is invariant under the Lorentz transformations. This is indeed so for plane waves and their localized superpositions. However, this is not the case for *evanescent* waves, which are well-defined only in a half-space, and are characterized by complex wave vectors. Here we describe transformations of evanescent electromagnetic waves and their polarization/spin/helicity properties under the Lorentz boosts along the three spatial directions.

## 1. Introduction

Vector properties of electromagnetic waves are intimately related to their relativistic nature and to the properties of photons, which are massless spin-1 particles. For a plane electromagnetic wave, the right-hand and left-hand circular polarizations correspond to the two helicity states of photons with the spin parallel and anti-parallel to the propagation direction. However, this simple relation between the polarization, spin, and helicity breaks down in the case of generic *inhomogeneous* electromagnetic fields, where these concepts correspond to essentially different physical properties.

Assuming a monochromatic field, the polarization is described by ellipses traced by the electric and magnetic fields at each point in space [1,2]. In turn, the spin angular momentum density is determined by the ellipticities and normals to these polarization ellipses [2,3]. At the same time, the helicity density turns out to be an independent pseudo-scalar quantity, which is not directly related to the geometry of the polarization ellipses [4–7]. Fundamentally, the spin angular momentum originates from the rotational symmetry of the space-time (i.e., one of the Poincaré-group symmetries) [7], while the helicity is related to the *dual symmetry* between the electric and magnetic fields [6–12], which is unrelated to the space-time properties. Therefore, it is usually assumed that the helicity of electromagnetic waves (and, generally, of massless particles) is a Lorentz invariant.

Recently, there was a great interest in unusual examples of the spin angular momentum in inhomogeneous optical fields [3,13–19], and in the role of the electromagnetic helicity in light-matter interactions [6,7,20–28]. These studies employed *evanescent* electromagnetic waves as simple and accessible examples of inhomogeneous fields with nontrivial spin and helicity properties [3,13,16–19,22,23]. However, most optical studies naturally assume a single laboratory reference frame and do not involve relativistic transformations.

Here we consider the Lorentz transformations of inhomogeneous electromagnetic waves, and show that evanescent waves exhibit unusual relativistic properties. Namely, in contrast to the common belief, the helicity is *not* conserved under the Lorentz boosts of evanescent waves. In particular, a transverse Lorentz boost of a linearly-polarized (zero-helicity) evanescent wave induces an elliptical polarization in the transverse plane and the corresponding non-zero helicity with the sign determined by the direction of the boost. This somewhat resembles the spin Hall effect of light and can be regarded as a relativistic example of the spin-orbit interactions of light [17] (see also Refs. [29,30] for nontrivial relativistic transformations of the optical angular momentum).

Two papers [31,32] with similar result came to my attention after the completion of this work. However, those papers did not use proper definitions of the spin and helicity densities, which were introduced only in recent years [3,6,7].

## 2. Basic equations

*2.1. Maxwell plane waves.*

We consider plane-wave-like monochromatic solutions of the Maxwell equations in free space. The electric and magnetic fields of these solutions can be written as $\mathbf{E}(\mathbf{r},t) = \text{Re}\left[\mathbf{E}_0 \exp(i\mathbf{k}\cdot\mathbf{r} - i\omega t)\right]$ and $\mathbf{H}(\mathbf{r},t) = \text{Re}\left[\mathbf{H}_0 \exp(i\mathbf{k}\cdot\mathbf{r} - i\omega t)\right]$. In what follows, all relations are linear, and we omit the real-part operation "Re", considering *complex* fields. The Maxwell equations take the form:

$$\mathbf{k}\cdot\mathbf{E} = \mathbf{k}\cdot\mathbf{H} = 0, \quad \mathbf{k}\times\mathbf{E} = \omega\mathbf{H}, \quad \mathbf{k}\times\mathbf{H} = -\omega\mathbf{E}. \tag{2.1}$$

Hereafter, we use natural Gaussian units with $c = 1$, and allow the wavevector $\mathbf{k}$ and frequency $\omega$ to take *complex* values (so that $\omega^2 = \mathbf{k}^2$ but generally $|\omega|^2 \neq |\mathbf{k}|^2$). Note also that we imply solutions with positive frequencies: $\text{Re}\,\omega > 0$; negative-frequency solutions are equivalent to their complex-conjugate with positive frequencies.

The *plane-wave* solution of Eqs. (2.1), propagating along the $z$-axis, can be written as

$$\mathbf{E} = A\frac{\hat{\mathbf{x}} + m\hat{\mathbf{y}}}{\sqrt{1+|m|^2}}\exp(ikz - i\omega t), \quad \mathbf{H} = \frac{\mathbf{k}}{k}\times\mathbf{E} = A\frac{-m\hat{\mathbf{x}} + \hat{\mathbf{y}}}{\sqrt{1+|m|^2}}\exp(ikz - i\omega t). \tag{2.2}$$

Here $A$ is the wave amplitude, the hats mark the unit vectors in the corresponding directions, $\mathbf{k} = k\hat{\mathbf{z}}$ is the real wavevector, and $m$ is a complex number which determines the polarization state of the wave [13,14,33] ($m = 0$ and $m = \infty$ correspond to the $x$- and $y$-linearly polarized waves, $m = \pm i$ correspond to the right- and left-hand circularly polarized waves, etc.).

The polarization parameter $m$ determines the spin and helicity of the plane wave. The cycle-averaged *spin angular momentum* density "per photon" (in units with $\hbar = 1$) in the wave (2.2) is given by [3,6,7,13–16]:

$$\mathbf{S} = \frac{1}{2N}\text{Im}\left(\mathbf{E}^*\times\mathbf{E} + \mathbf{H}^*\times\mathbf{H}\right) = \frac{2\,\text{Im}\,m}{1+|m|^2}\hat{\mathbf{z}}, \tag{2.3}$$

where $N = \left(|\mathbf{E}|^2 + |\mathbf{H}|^2\right)/2 = |A|^2$ is the normalization factor. In turn, the *helicity* density "per photon" is determined by [3,6,7,20,21,28]:

$$\mathfrak{S} = \frac{1}{N}\text{Im}\left(\mathbf{H}^*\cdot\mathbf{E}\right) = \frac{2\,\text{Im}\,m}{1+|m|^2}. \tag{2.4}$$

The right-hand sides of Eqs. (2.3) and (2.4) reveal the close relation between the spin and helicity of plane waves: $\mathfrak{S} = \mathbf{S}\cdot\mathbf{k}/k$.

*2.2. Evanescent waves.*

Consider now *evanescent waves*, which can appear near planar interfaces in a variety of optical systems [34]. Assuming propagation along the $z$-axis and exponential decay in the $x$-direction (so that the solution is well-defined in the $x > 0$ half-space), the evanescent wave can be presented as a plane-wave-like solution with the *complex wave vector*: $\mathbf{k} = k_z\hat{\mathbf{z}} + i\kappa\hat{\mathbf{x}}$, $\mathbf{k}^2 = k_z^2 - \kappa^2 = \omega^2 < k_z^2$. The normalized electric and magnetic fields of this wave are [3,13]:

$$\mathbf{E} = \frac{\hat{\mathbf{x}} + m\frac{k}{k_z}\hat{\mathbf{y}} - i\frac{\kappa}{k_z}\hat{\mathbf{z}}}{\sqrt{1+|m|^2}} \exp(ik_z z - \kappa x - i\omega t),$$

$$\mathbf{H} = \frac{\mathbf{k}}{k} \times \mathbf{E} = \frac{-m\hat{\mathbf{x}} + \frac{k}{k_z}\hat{\mathbf{y}} + im\frac{\kappa}{k_z}\hat{\mathbf{z}}}{\sqrt{1+|m|^2}} \exp(ik_z z - \kappa x - i\omega t). \quad (2.5)$$

Here and in what follows, we imply $k \equiv \omega < k_z$.

The spin and helicity properties of evanescent waves (2.5) are less trivial than those of plane waves, Eqs. (2.2)–(2.4). The normalized spin and helicity densities (2.3) and (2.4) [with $N = |A|^2 \exp(-2\kappa x)$] read [3,13]:

$$\mathbf{S} = \frac{2\,\mathrm{Im}\,m}{1+|m|^2}\frac{k}{k_z}\hat{\mathbf{z}} + \frac{\kappa}{k_z}\hat{\mathbf{y}}, \quad \mathfrak{S} = \frac{2\,\mathrm{Im}\,m}{1+|m|^2}. \quad (2.6)$$

The second term in the first equation (2.6) describes the *transverse helicity-independent spin density* of evanescent fields, which is currently attracting considerable attention [3,13–19]. Interestingly, the simple plane-wave equation for the helicity, $\mathfrak{S} = \mathbf{S} \cdot \mathbf{k}/k$, holds true for evanescent waves.

*2.3. Lorentz boosts.*

We will deal with the Lorentz transformations (boosts) along the Cartesian axes of the problem. For example, a longitudinal $z$-boost with the velocity $\mathbf{v} = v\hat{\mathbf{z}}$ yields the transformation of the space-time in the moving reference frame [35]:

$$t' = \gamma(t - vx), \quad x' = x, \quad y' = y, \quad z' = \gamma(z - vt), \quad (2.7)$$

where $\gamma = 1/\sqrt{1-v^2}$ is the Lorentz factor. Since the wave phase $\mathbf{k}\cdot\mathbf{r} - \omega t$ is a Lorentz scalar, the four-wavevector (even a complex one) is transformed in a similar manner:

$$\omega' = \gamma(\omega - vk_z), \quad k'_x = k_x, \quad k'_y = k_y, \quad k'_z = \gamma(k_z - v\omega). \quad (2.8)$$

The electric and magnetic field amplitudes, i.e., the pre-exponential factors in Eqs. (2.2) and (2.5), are transformed as components of the rank-2 field tensor [35]:

$$E'_x = \gamma(E_x - vH_y), \quad E'_y = \gamma(E_y + vH_x), \quad E'_z = E_z,$$

$$H'_x = \gamma(H_x + vE_y), \quad H'_y = \gamma(H_y - vE_x), \quad H'_z = H_z, \quad (2.9)$$

The Lorentz transformations under the boosts along the $x$ and $y$ axes are obtained from Eqs. (2.7)–(2.9) with the corresponding permutation of indices.

*2.4. Lorentz boosts of plane waves.*

As the first simple example, we consider a longitudinal $z$-boost of the plane wave (2.2). Using Eqs. (2.7)–(2.9), we readily find that the transformed wave has the same form (2.2) of a $z$-propagating plane wave:

$$\mathbf{E}' = A'\frac{\hat{\mathbf{x}}' + m\hat{\mathbf{y}}'}{\sqrt{1+|m|^2}}\exp(ik'z' - i\omega't'), \quad \mathbf{H}' = \frac{\mathbf{k}'}{k'} \times \mathbf{E}'. \quad (2.10)$$

Here, $\omega' = \gamma(\omega - vk)$, $\mathbf{k}' = \gamma(k - v\omega)\hat{\mathbf{z}} \equiv k'\hat{\mathbf{z}}$, and $A' = (k'/k)A$. Importantly, the polarization parameter $m$, and, hence, the spin (2.3) and helicity (2.4), remain unchanged in the boosted wave (2.10): $\mathbf{S}' = \mathbf{S}$ and $\mathfrak{S}' = \mathfrak{S}$, Fig. 1a.

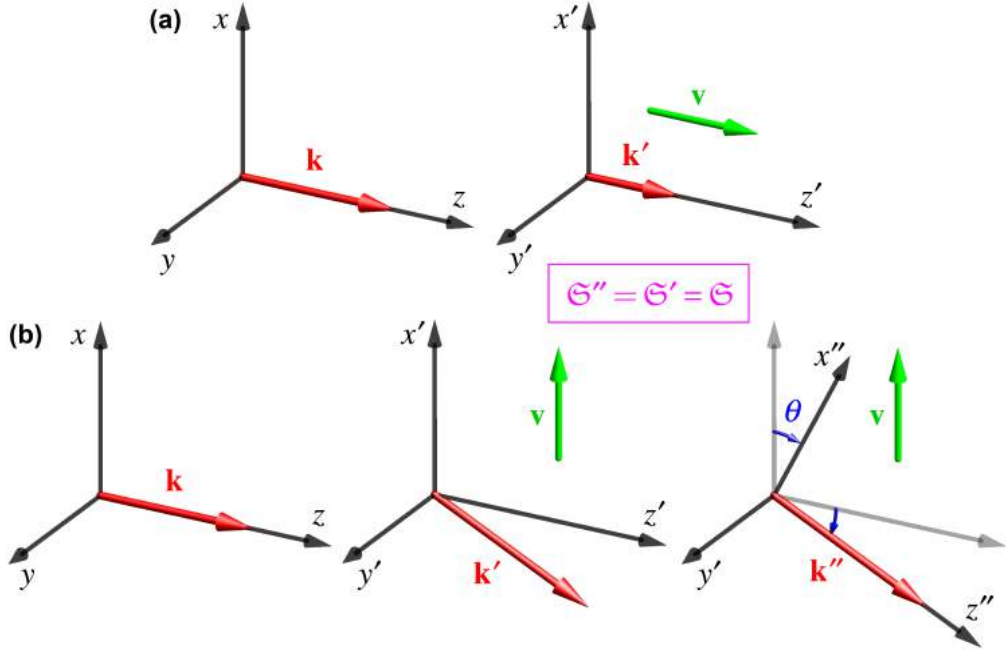

**Fig. 1.** (a) The longitudinal Lorentz boost of a plane wave, Eq. (2.10). (b) The transverse Lorentz boost and aligning rotation of a plane wave, Eqs. (2.11) and (2.12). The plane-wave helicity $\mathfrak{S}$ is invariant under the Lorentz transformations.

Consider now a transverse Lorentz boost (e.g., along the $x$-axis, $\mathbf{v} = v\hat{\mathbf{x}}$) of the plane wave (2.2). The transformed wave will propagate obliquely in the $(x',z')$-plane:

$$\mathbf{E}' = A\frac{\hat{\mathbf{x}}' + \gamma m\hat{\mathbf{y}}' + \gamma v\hat{\mathbf{z}}'}{\sqrt{1+|m|^2}}\exp(i\mathbf{k}'\cdot\mathbf{r}' - i\omega't'), \quad \mathbf{H}' = \frac{\mathbf{k}'}{k'}\times\mathbf{E}', \qquad (2.11)$$

where $\omega' = \gamma\omega$ and $\mathbf{k}' = -\gamma vk\hat{\mathbf{x}}' + k\hat{\mathbf{z}}'$, Fig. 1b. To compare the boosted wave (2.11) with the $z$-propagating wave (2.2), it is instructive to rotate the coordinates of the boosted reference frame in the $(z',x')$-plane in order to align the rotated $z''$-axis with the wave propagation direction [30]. Performing rotation $(x',z') \to (x'',z'')$ of Eq. (2.11) by the angle $-\theta$: $\sin\theta = v$, $\cos\theta = 1/\gamma$, Fig. 1b, we arrive at

$$\mathbf{E}'' = A'\frac{\hat{\mathbf{x}}'' + m\hat{\mathbf{y}}'}{\sqrt{1+|m|^2}}\exp(ik'z'' - i\omega't'), \quad \mathbf{H}'' = \frac{\mathbf{k}''}{k'}\times\mathbf{E}'', \qquad (2.12)$$

where $A' = (k'/k)A$ and $\mathbf{k}'' = k'\hat{\mathbf{z}}''$. Akin to Eq. (2.10), the polarization parameter $m$, and, hence, the spin/helicity properties (2.3) and (2.4), are preserved in the transversely boosted wave (2.12):

$$\mathbf{S}'' = \frac{2\operatorname{Im} m}{1+|m|^2}\hat{\mathbf{z}}'', \quad \mathfrak{S}'' = \mathfrak{S}' = \frac{2\operatorname{Im} m}{1+|m|^2}. \qquad (2.13)$$

Thus, we conclude that the Lorentz transformations do not change the spin and helicity of plane electromagnetic waves (apart from the rotations of the propagation/spin direction: $\hat{\mathbf{z}} \to \hat{\mathbf{z}}''$). In particular, the helicity is a Lorentz pseudo-scalar, i.e., is invariant under the Lorentz transformations [4–12].

## 3. Lorentz boosts of evanescent waves

We now examine the main subject of this work, namely, the Lorentz transformations of evanescent electromagnetic waves. An evanescent wave (2.5) is a strongly anisotropic entity, and therefore we consider all three Lorentz boosts along the three Cartesian axes. These transformations produce qualitatively different modifications of the boosted fields.

*3.1. Longitudinal boost.*

We start with the simplest case of a longitudinal Lorentz boost (2.7)–(2.9) of the field (2.5) along the propagation $z$-axis. This results in a similar evanescent wave with modified parameters:

$$\mathbf{E}' = A' \frac{\hat{\mathbf{x}}' + m \frac{k'}{k'_z} \hat{\mathbf{y}}' - im \frac{\kappa'}{k'_z} \hat{\mathbf{z}}'}{\sqrt{1+|m|^2}} \exp\left(ik'_z z' - \kappa' x' - i\omega' t'\right), \quad \mathbf{H}' = \frac{\mathbf{k}'}{k'} \times \mathbf{E}', \quad (3.1)$$

where $\mathbf{k}' = k'_z \hat{\mathbf{z}}' + i\kappa' \hat{\mathbf{x}}'$, $k'_z = \gamma(k_z - vk)$, $\kappa' = \kappa$, $\omega' = \gamma(k - vk_z)$, and $A' = (k'_z/k_z)A$.

At first glance, the polarization parameter $m$, and hence the helicity $\mathfrak{S}$, are unaffected by the longitudinal boost. At the same time, the spin density (2.6) of the boosted wave (3.1) changes because it depends not only on $m$, but also on the wave-vector components:

$$\mathbf{S}' = \frac{2\,\mathrm{Im}\,m}{1+|m|^2} \frac{k'}{k'_z} \hat{\mathbf{z}}' + \frac{\kappa'}{k'_z} \hat{\mathbf{y}}' \neq \mathbf{S}, \quad \mathfrak{S}' = \mathfrak{S}. \quad (3.2)$$

However, remarkably, there is a critical velocity $v_c < 1$, for which the frequency of the evanescent wave *vanishes*, becoming *negative* for larger than critical velocities:

$$\omega' \leq 0 \quad \text{for} \quad v \geq v_c = k/k_z. \quad (3.3)$$

Since only the real part of the complex field (3.1) makes physical sense, we should take the complex conjugate of the negative-frequency solution, transforming it to the equivalent positive-frequency solution for $v > v_c$:

$$\mathbf{E}'' = \mathbf{E}'^* = A'' \frac{\hat{\mathbf{x}}' + m'' \frac{k''}{k''_z} \hat{\mathbf{y}}' - im'' \frac{\kappa'}{k''_z} \hat{\mathbf{z}}'}{\sqrt{1+|m|^2}} \exp\left(ik''_z z' - \kappa' x' - i\omega'' t'\right), \quad \mathbf{H}'' = \frac{\mathbf{k}''}{k''} \times \mathbf{E}'', \quad (3.4)$$

This evanescent wave propagates in the *negative-$z$* direction with $\mathbf{k}'' = k''_z \hat{\mathbf{z}}' + i\kappa' \hat{\mathbf{x}}'$, $k''_z = -k'_z < 0$, $k'' = \omega'' = -\omega' > 0$, and $A'' = A'^*$, Fig. 2. Most significantly, the complex conjugation also changes the polarization parameter: $m'' = m^*$. This *reverses* the handedness of the wave, together with its *spin* and *helicity*:

$$\mathbf{S}'' = -\frac{2\,\mathrm{Im}\,m}{1+|m|^2} \frac{k'}{k'_z} \hat{\mathbf{z}}' - \frac{\kappa'}{k'_z} \hat{\mathbf{y}}' = -\mathbf{S}', \quad \mathfrak{S}'' = -\frac{2\,\mathrm{Im}\,m}{1+|m|^2} = -\mathfrak{S}. \quad (3.5)$$

Thus, in contrast to the intuition based on plane waves and localized solutions, *helicity of evanescent waves is not invariant under Lorentz transformations*. This is in agreement with earlier calculations of [31]. It should be noticed that despite the formal difference in signs between Eqs. (3.2) and (3.5), only the transverse component of the spin, $S_y$, and helicity $\mathfrak{S}$ flip upon the transition through the critical velocity. At the same time, the longitudinal spin component $S_z$ preserves its sign. This is because $k' > 0$ in Eq. (3.2) but $k' < 0$ in Eq. (3.5).

Thus, the plane-wave helicity relation $\mathfrak{S}'' = \mathbf{S}'' \cdot \mathbf{k}''/k''$ is fulfilled, and the helicity (3.5) flips because of the reversed propagation direction with the same longitudinal spin.

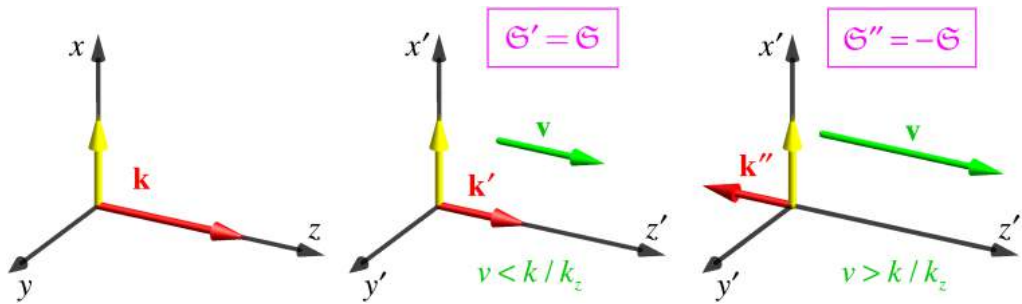

**Fig. 2.** The longitudinal Lorentz boosts of the evanescent wave, Eqs. (2.5), (3.1), and (3.4). Here and in the figures below, the real and imaginary parts of complex wave vectors are shown by the red and yellow arrows, respectively. The boosts with over-critical velocities, higher than the velocity of the evanescent-wave propagation, reverse the real part of the wave vector, as well as the helicity and the transverse spin component of the wave, Eqs. (3.5).

The above results have a clear physical explanation. Indeed, the evanescent wave (2.5) is characterized by *subluminal* phase and group velocities: $v_{wave} = k/k_z < 1$ [19,36]. Therefore, a longitudinal Lorentz boost with a larger velocity $v > v_{wave}$ reverses the direction of the wave propagation: $k_z'' < 0$. However, this does not change the sense of the electric field rotation in the transverse $(x,y)$-plane (the longitudinal $z$-component of the spin does not change its sign), and therefore effectively reverses the wave helicity determined with respect to the inverted longitudinal wavevector component. Note that the transverse spin component $S_y$ flips due to the "transverse spin-momentum locking" [3,16–18,37]. It is also worth remarking that the longitudinal wavevector component never vanishes: it reaches the minimal positive value $k_{z\min}' = \gamma \kappa^2 / k_z$ at $v = v_{wave} - 0$ and then jumps to the maximum negative value: $k_{z\max}' = -\gamma \kappa^2 / k_z$ for $v = v_{wave} + 0$. One of the observable manifestations of the anomalous behavior of evanescent waves near $v = v_{wave}$ is the resonant interaction of an electron moving along a metal-vacuum interface with surface plasmon-polaritons (which have the evanescent-wave form in vacuum), when the electron velocity coincides with that of the surface wave [38,39].

*3.2. Vertical boost.*

Consider now the Lorentz boost of the evanescent wave (2.5) in the vertical $x$-direction: $\mathbf{v} = v\hat{\mathbf{x}}$ (we assume the interface $x = 0$ to be horizontal). This produces the following transformation of the complex wave vector: $\mathbf{k}' = k_z'\hat{\mathbf{z}}' + k_x'\hat{\mathbf{x}}'$, $k_z' = k_z$, $k_x' = \gamma(i\kappa - vk)$, and frequency $\omega' = \gamma(k - iv\kappa) = k'$, Fig. 3. These parameters correspond to a *spatio-temporal* evanescent wave decaying both in space and time (for other examples of spatio-temporal inhomogeneous waves, see [40–42]). This wave propagates obliquely in the $(z,x)$-plane. Therefore, akin to Section 2.4, we perform an additional rotation of this plane by the angle $-\theta$: $\sin\theta = \gamma vk / \sqrt{\gamma^2 k^2 + \kappa^2}$, $\cos\theta = k_z / \sqrt{\gamma^2 k^2 + \kappa^2}$, to align the rotated $z''$-axis with the wave propagation $\mathrm{Re}\,\mathbf{k}'$, Fig. 3. This results in the wave vector $\mathbf{k}'' = k_z''\hat{\mathbf{z}}'' + k_x''\hat{\mathbf{x}}''$ with $k_z'' = \sqrt{\gamma^2 k^2 + \kappa^2} - i\gamma^2 v\kappa k / \sqrt{\gamma^2 k^2 + \kappa^2}$ and $k_x'' = i\gamma \kappa k_z / \sqrt{\gamma^2 k^2 + \kappa^2} \equiv i\kappa''$.

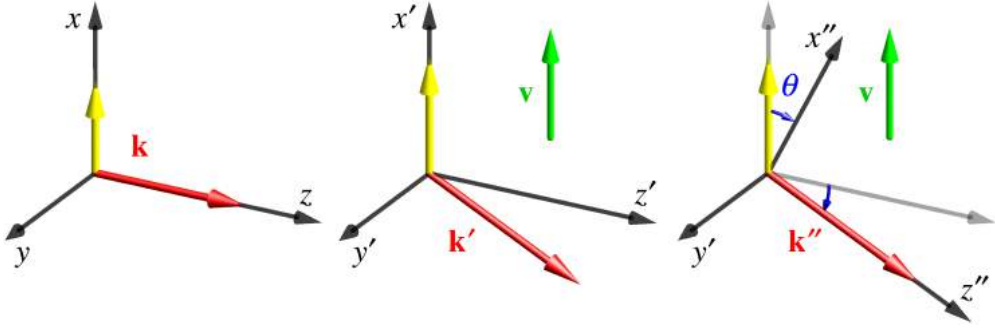

**Fig. 3.** Vertical Lorentz boost and aligning rotation of the evanescent wave, Eqs. (3.6). The real and imaginary parts of the boosted wave vector are not orthogonal anymore, which indicates the *spatio-temporal* (complex-frequency) character of the boosted wave.

The wave field, after the above boost and rotation, takes the following form:

$$\mathbf{E}'' = A'' \frac{\hat{\mathbf{x}}'' + m\frac{k'}{k_z''}\hat{\mathbf{y}}' - im\frac{\kappa''}{k_z''}\hat{\mathbf{z}}''}{\sqrt{1+|m|^2}} \exp(ik_z''z'' - \kappa''x'' - i\omega't'), \quad \mathbf{H}'' = \frac{\mathbf{k}''}{k'} \times \mathbf{E}'', \qquad (3.6)$$

where $A'' = (k_z''/k_z)A$. The spatio-temporal evanescent wave (3.6) decays in time ($\operatorname{Im}\omega' < 0$) when $\kappa v > 0$, i.e., the boosted reference frame moves away from the interface in the direction of the spatial decay of the original wave (2.5). Notably, the polarization parameter $m$ remains unchanged after the vertical Lorentz boost. This suggests that the helicity is also unaffected by this Lorentz transformation. At the same time, the spin density is changed because of the changed wave-vector components. However, it should be emphasized that Eqs. (2.6) are *not* applicable anymore, because these were derived in the assumption of a *real* frequency and longitudinal wave-vector component. Rigorous calculation of the helicity and spin of the spatio-temporal evanescent wave (3.6) requires extension of these properties to the case of complex-frequency waves, which is beyond the scope of the present work. We only note that the complex $k'$ and $k_z''$ in Eq. (3.6) result in the *elliptical* polarization (rotating electric field) in the transverse $(x'',y')$ plane even at $\operatorname{Im} m = 0$ (which corresponds to linear polarizations of real-frequency waves). Therefore, the usual intuition about spin and helicity properties does not work in this case.

*3.3. Transverse boost.*

We finally examine the remaining Lorentz boost of the evanescent wave (2.5) in the *transverse* $y$-direction: $\mathbf{v} = v\hat{\mathbf{y}}$, Fig. 4. It produces the wave vector $\mathbf{k}' = i\kappa\hat{\mathbf{x}}' + k_y'\hat{\mathbf{y}}' + k_z\hat{\mathbf{z}}'$ with $k_y' = -\gamma v k$, and the frequency $\omega' = \gamma\omega = k'$. Performing rotation of the $(y',z')$-plane by the angle $\theta$: $\sin\theta = \gamma v k/\sqrt{\gamma^2 k^2 + \kappa^2}$, $\cos\theta = k_z/\sqrt{\gamma^2 k^2 + \kappa^2}$, we obtain the wave vector $\mathbf{k}'' = i\kappa\hat{\mathbf{x}}' + k_z''\hat{\mathbf{z}}''$ with $k_z'' = \sqrt{\gamma^2 k^2 + \kappa^2}$. This is a usual evanescent wave of the form (2.5) propagating in the $z''$-direction and decaying in the $x$-direction.

The corresponding transformed wave field becomes:

$$\mathbf{E}'' = A'' \frac{\hat{\mathbf{x}}' + m''\frac{k'}{k_z''}\hat{\mathbf{y}}'' - im''\frac{\kappa}{k_z''}\hat{\mathbf{z}}''}{\sqrt{1+|m|^2}} \exp(ik_z''z'' - \kappa x' - i\omega't'), \quad \mathbf{H}'' = \frac{\mathbf{k}''}{k'} \times \mathbf{E}''. \qquad (3.7)$$

where

$$A'' = \frac{\gamma(k_z + imv\kappa)}{k_z}A, \quad m'' = \frac{mk_z - iv\kappa}{k_z + imv\kappa}. \tag{3.8}$$

These equations represent interesting results. First, the wave amplitude is modified depending on the ellipticity of the polarization of the initial wave and the direction of the boost. In particular, the right-hand ($m = i$) and left-hand ($m = -i$) circularly-polarized waves have higher amplitudes after the boosts with $v < 0$ and $v > 0$, respectively. Second, the polarization parameter $m$ is modified by the transverse boost, and the initial linearly-polarized waves (e.g., $m = 0$ or $m = \infty$) become right-hand and left-hand elliptically polarized after the boosts with $v < 0$ and $v > 0$, respectively. This is in agreement with earlier calculations in [32].

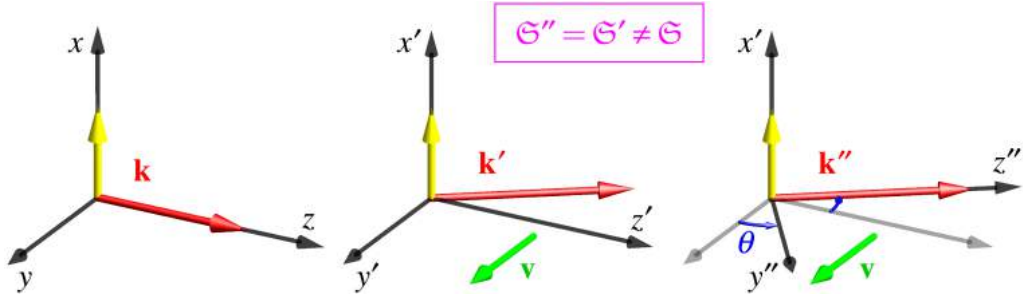

**Fig. 4.** Transverse Lorentz boost and aligning rotation of the evanescent wave, Eq. (3.7). This boost essentially changes the polarization, helicity, and spin of the wave, Eqs. (3.8) and (3.9). In particular, a linearly-polarized (zero-helicity) wave acquires a non-zero boost-induced helicity.

Importantly, the modification of the polarization parameter (3.8) changes both the spin and helicity densities (2.6) in the transversely-boosted evanescent wave:

$$\mathbf{S}'' = \frac{2\operatorname{Im} m''}{1+|m''|^2}\frac{k'_z}{k''_z}\hat{\mathbf{z}}'' + \frac{\kappa}{k''_z}\hat{\mathbf{y}}'', \quad \mathfrak{S}'' = \frac{2\operatorname{Im} m''}{1+|m''|^2} \neq \mathfrak{S}. \tag{3.9}$$

In particular, for the $x$-linearly polarized initial wave, $m = 0$, $m'' = -iv\kappa/k_z$, and the helicity density of the boosted wave becomes $\mathfrak{S}'' = -\dfrac{2v\kappa/k_z}{1+(v\kappa/k_z)^2}$. For the $y$-polarized wave, $m = \infty$, we obtain $m'' = -ik_z/v\kappa$, and the helicity density $\mathfrak{S}'' = -\dfrac{2k_z/v\kappa}{1+(k_z/v\kappa)^2}$. This confirms the conclusion of the Section 3.1 that *the helicity of evanescent waves is not Lorentz-invariant*. It is worth remarking that the Lorentz-boost-induced helicity $\mathfrak{S}''$ involves combinations of the main three parameters of the problem: $k_z$, $\kappa$, and $v$. Their product is a pseudo-scalar quantity (*P*-odd and *T*-even) because $k_z$ and $v$ are *P*-odd and *T*-odd quantities, while $\kappa$ is a *P*-odd and *T*-even quantity [3]. Note that the plane-wave helicity relation $\mathfrak{S}'' = \mathbf{S}'' \cdot \mathbf{k}''/k''$ still holds true.

The symmetries and geometry of the nontrivial polarization properties of the Lorentz-boosted evanescent waves allow us to consider them as novel manifestations of the *spin-orbit interactions of light* [17]. Indeed, the helicity-dependent enhancement of the wave amplitude for the wave boosted in the transverse direction corresponds to the *spin Hall effect of light*. In particular, it is observed for surface plasmon-polaritons (involving evanescent waves), where the helicity of the incident wave generates a helicity-dependent transverse wave-vector component of the scattered wave [43]. This component can be associated with the *motion of the reference frame* in the transverse direction. Another phenomenon with similar geometry is the helicity-dependent transverse *drag effect* for surface plasmon-polaritons [44,45]. There, the incident wave induces a helicity-dependent transverse current in the metal, i.e., the helicity-dependent transverse motion of electrons in the metal. In turn, the Lorentz-boost-induced helicity $\mathfrak{S}''$ can be considered as *inversion* of the spin-Hall effect because it involves the transverse-transport-induced helicity instead of the helicity-induced transverse transport.

## 4. Discussion

Inhomogeneous electromagnetic waves are ubiquitous in modern optics and photonics. Their properties cannot be described by the simplified plane-wave model, and one should carefully distinguish between the notions of polarization, spin, and helicity [2,3,6,7,46]. Moreover, the properties of localized and delocalized solutions can also differ considerably. For example, electric and magnetic contributions to the integral spin angular momentum of a localized electromagnetic field are equal to each other [47], but this is not the case for unbounded evanescent waves [3,13].

Evanescent waves represent mathematically simple and physically nontrivial (inhomogeneous and unbounded) electromagnetic waves. Examples considered in this work show that the Lorentz invariance of the helicity (proven for plane waves and their localized superpositions) breaks down for evanescent waves. Of course, evanescent waves are well-defined only in *half-space*, and all quantities can be defined either as local densities or with a half-space integration. For any properly localized wave (a beam or a wave packet), the integral helicity should be invariant under the transverse Lorentz boosts. Nonetheless, transformations of the helicity in the evanescent waves hint that the *local* helicity density in a beam or a wave packet can be affected by the Lorentz transformations [30].

It is worth noticing that the Lorentz transformations of the spin of relativistic particles (including photons) are described by the *Pauli-Lubanski four-vector* [48–50]. For plane waves with the momentum $\mathbf{k}$ and energy $\omega$, it can be defined as $\Sigma^\mu \equiv (\Sigma_0, \boldsymbol{\Sigma}) = (\mathbf{S}\cdot\mathbf{k}, \mathbf{S}\omega)$. It is easy to see that this quantity is indeed transformed as a four-vector under the Lorentz boosts of plane waves, Section 2.4. Furthermore, transformations of the integral spin and helicity of localized wavepacket states are described by a similar Pauli-Lubanski vector involving integral quantities [50]. However, for evanescent waves with complex wavevectors, the quantity $(\mathbf{S}\cdot\mathbf{k}, \mathbf{S}\omega)$ does *not* follow the four-vector transformations; this can be readily seen from the equations of Section 3. This confirms once again that evanescent waves, possessing abnormal properties, such as the transverse spin [3,13–19], cannot be regarded as the "usual" photon states in the Hilbert space. Nonetheless, the simple plane-wave relation for the helicity, $\mathfrak{S} = \mathbf{S}\cdot\mathbf{k}/k$, holds true under the Lorentz boosts of evanescent waves.

Despite being defined only over half-space, evanescent waves are important physical entities observable in many experimental situations [34]. Moreover, the Lorentz transformations of evanescent waves naturally appear in problems involving moving charged particles. Indeed, the electromagnetic field generated by a uniformly moving charged particle can be expanded into a set of evanescent waves in any half-space not including the particle path [51]. This underpins various Cherenkov-type effects for electrons interacting with surface waves and gratings [38,29,51,52]. The results of this work can help in understanding and predicting the helicity- and spin-dependent aspects of such phenomena by performing the Lorentz transformations between the laboratory frame (with evanescent waves attached to the interface) and the electron rest frame.

## Acknowledgements

I acknowledge helpful discussions with Y. P. Bliokh, E. A. Ostrovskaya, and G. Molina-Terriza. This work was supported by the Australian Research Council.